\documentclass[runningheads]{llncs}
\usepackage[T1]{fontenc}
\usepackage{graphicx}
\usepackage{amsmath}
\usepackage{cite}
\usepackage{amsmath,amssymb,amsfonts}
\usepackage{textcomp}
\usepackage{caption}
\usepackage{subfigure}
\usepackage{multirow}
\usepackage{multicol}
\usepackage{booktabs}

\usepackage[noend]{algpseudocode}
\usepackage{algorithmicx,algorithm}
\usepackage{verbatim}
\usepackage{array}
\usepackage{enumerate}
\usepackage{enumitem}

\begin{document}
\title{Quantum Meet-in-the-Middle Attack on Feistel Construction}
%
%
\author{Yinsong, Xu\inst{1,2,3}\orcidID{0000-0002-0585-318X} \and
Zheng, Yuan\inst{1,2,3,*} }
\authorrunning{Y. Xu et al.}
%
\institute{State Key Laboratory of Networking and Switching Technology, Beijing University of Posts and Telecommunications, Beijing, 100876, China. \and
Beijing Electronic Science and Technology Institute, Beijing, 100070, China. \and
Advanced Cryptography and System Security Key Laboratory of Sichuan Province, Chengdu, 610103, China.\\
* Corresponding Author\\
\email{mugongxys@foxmail.com,zyuan@tsinghua.edu.cn}\\
}
\maketitle              
\begin{abstract}
Inspired by Hosoyamada \textit{et al.}'s work \cite{Hosoyamada2018A}, we propose a new quantum meet-in-the-middle (QMITM) attack on $r$-round ($r \ge 7$) Feistel construction to reduce the time complexity. Similar to Hosoyamada \textit{et al.}'s work, our attack on 7-round Feistel is also based on Guo \textit{et al.}'s classical meet-in-the-middle (MITM) attack \cite{Guo2016}. The classic MITM attack consumes a lot of time mainly in three aspects: construct the lookup table, query data and find a match. Therefore, parallel Grover search processors are used to reduce the time of constructing the lookup table. And we adjust the truncated differentials of the 5-round distinguisher proposed by Guo \textit{et al.} to balance the complexities between constructing the lookup table and querying data. Finally, we introduce a quantum claw finding algorithm to find a match for reducing time. The subkeys can be recovered by this match.  Furthermore, for $r$-round ($r > 7$) Feistel construction, we treat the above attack on the first 7 rounds as an inner loop and use Grover's algorithm to search the last $r-7$ rounds of subkeys as an outer loop. In summary, the total time complexity of our attack on $r$-round ($r \ge 7$) is only $O(2^{2n/3+(r-7)n/4})$ less than classical and quantum attacks. Moreover, our attack belongs to Q1 model and is more practical than other quantum attacks.

\keywords{Quantum meet-in-the-middle attack \and Feistel construction \and Quanutm claw finding algorithm \and Grover's algorithm.}
\end{abstract}
\section{Introduction}\label{sec_1}
A Feistel network is a scheme that builds $n$-bit permutations from smaller, usually $n/2$-bit permutations or functions \cite{Feistel1975}. The Feistel-based design approach is widely used in block ciphers. In particular, a number of current and former international or national block cipher standards such as DES \cite{Coppersmith1994}, Triple-DES \cite{ISO2010} and Camellia \cite{Aoki2009} are Feistel ciphers. 

\begin{figure}[htbp]
	\centering
	\includegraphics[width=3cm]{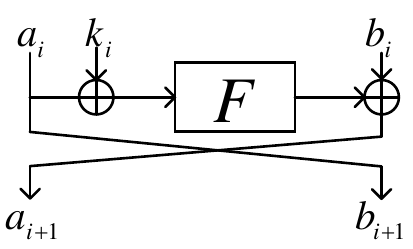}
	\caption{Our target Feistel construction.}
	\label{fig_feistel_construction}
\end{figure}

Feistel ciphers have many constructions, and the analyzed target construction in this paper is displayed in Fig. \ref{fig_feistel_construction}. An $n$-bit state is divided into $n/2$-bit halves denoted by ${a_i}$ and ${b_i}$, and the state is updated by iteratively applying the following two operations:
\begin{equation}\nonumber
	{a_{i + 1}} \leftarrow {a_i} \oplus F({k_i} \oplus {b_i}), \ {b_{i + 1}} \leftarrow {a_i},
\end{equation}
where $F$ is a public function and $k_i$ is a subkey with $n/2$ bits. Note that, the target Feistel construction is also called Feistel-2 in Ref. \cite{Guo2016}, or called Feistel-KF in Ref. \cite{Ito2019}. For brevity, we simply call it the Feistel construction.

For the target Feistel construction, Guo \textit{et al.} \cite{Guo2016} absorbed the idea of meet-in-the-middle attack proposed by Demirci and Sel\c cuk \cite{Demirci2008}, and applied it on the attack of the Feistel construction. They firstly computed all the possible sequences constructed from a $\delta$-set such that a pair of messages satisfy the proposed 5-round distinguisher. Then, they collected enough pairs of plaintexts and corresponding ciphertexts, and guessed the subkeys out of the distinguisher to compute the corresponding sequence for each pair. Finally, if the computed sequence from each pair of collected plaintexts belongs to the precomputed sequences from the distinguisher, the guessed subkeys are correct; otherwise, they are wrong. In their attack on 6-round Feistel construction, its time complexity is $O(2^{3n/4})$. Later, Zhao \textit{et al.} combined Guo \textit{et al.}'s work and pairs sieve procedure to attack on 7-round Feistel construction with $O(2^n)$ time. Besides, some other  attacks penetrate up to 6 rounds, such as impossible differentials \cite{Knudsen2002}, all-subkey recovery \cite{Isobe2013A,Isobe2013B} and integral-like attacks \cite{Todo2013}.

In the quantum setting, cryptanalysts hope to take advantage of quantum computing to further reduce attack complexity. For symmetric cryptosystems, Grover's algorithm \cite{Grover1997} can provide a quadratic speedup over exhaustive search on keys. In addition, some other quantum algorithms have been applied to the analysis of block ciphers and achieved good results, such as Simon's algorithm \cite{Simon1997} and Bernstein-Vazirani algorithm \cite{Bernstein1997}. Many block ciphers have been evaluated for the security in the quantum setting, e.g. against Even-Mansour cipher \cite{Kuwakado2012}, MACs\cite{Kaplan2016B,Santoli2016}, AEZ\cite{Shi2018}, AES-COPA\cite{Xu2021}, FX construction\cite{Leander2017} and so on. To evaluate the adversary's ability, Kaplan divides these quantum attacks into two attack models: Q1 model and Q2 model \cite{Kaplan2016A}.
\begin{enumerate}
	\item \textbf{Q1 model.} The adversary is allowed only to make classical online queries and performs quantum offline computation. 
	
	\item \textbf{Q2 model.} The adversary can make quantum superposition online queries for cryptographic oracle and performs quantum offline computation. 
\end{enumerate}
Obviously, the adversary is more practical in Q1 model, and more powerful in Q2 model.

Many researches have analyzed the security of the Feistel construction in Q1 or Q2 model. In 2010, Kuwakado \textit{et al.} \cite{Kuwakado2010} proposed a quantum 3-round Feistel distinguisher and used Simon's algorithm to recover subkeys in Q2 model. Its time complexity only needs $O(n)$ because of Simon's algorithm. Based on Kuwakado \textit{et al.}'s work, Dong \textit{et al.} \cite{Dong2018A,Dong2019A} used Grover's algorithm to search the last $r-3$ rounds subkeys of $r$-round Feistel construction. Its time complexity is $O({2^{0.25nr - 0.75n}})$. 

Different from Kuwakado \textit{et al.}'s quantum 3-round Feistel distinguisher, Xie \textit{et al.} \cite{Xie2019} and Ito \textit{et al.} \cite{Ito2019} proposed new quantum Feistel distinguishers respectively. Xie \textit{et al.} \cite{Xie2019} used Bernstein-Vazirani algorithm instead of Simon's algorithm to recover subkeys. However, this modification cause a slight increase in complexity with $O(n^2)$ time. Ito \textit{et al.} \cite{Ito2019} proposed a new 4-round Feistel distinguisher, and used Grover's algorithm to search the last $r-4$ rounds subkeys similarly to Dong \textit{et al.}'s attack. Besides, Dong \textit{et al.} \cite{Dong2020A} and Bonnetain \textit{et al.} \cite{Bonnetain2020} proposed quantum slide attacks on Feistel construction by Simon's algorithm respectively. 

The quantum attacks listed above are all Q2 models, Hosoyamada \textit{et al.} \cite{Hosoyamada2018A} proposed a new quantum attack on 6-round Feistel construction that belongs to Q1 model. They gave a quantum claw finding algorithm to find a match between two tables, where two tables are constructed by two phases of Guo \textit{et al.}'s attack: one table from precomputation and the other one from collected pairs. If there exists one match, the guessed subkeys are correct.

\noindent\textbf{Contribution.} Inspired by Hosoyamada \textit{et al.}'s work \cite{Hosoyamada2018A}, we firstly propose a new quantum meet-in-the-middle attack (QMITM) on 7-round Feistel construction to reduce time complexity in Q1 model. Similar to Guo \textit{et al.}'s attack, we divide our attack into two phases: pre-computation phase and queried-data analysis phase. In the pre-computation phase, we use parallel Grover search processors to compute all possible sequence through the 5-round distinguisher, which are stored in a classical lookup table. To balance the complexity between pre-computation phase and queried-data analysis phase, we adjust truncated differentials of the 5-round distinguisher. Then, we transform these classical queried-data to a superposition state by quantum random access memory, and compute corresponding sequence for each pair by quantum operations. Finally, we use quantum claw finding algorithm \cite{Buhrman2005} to find a claw between the table and the superposition state and recover subkeys through this found claw. In summary, the time complexity of our attack is only $O(2^{2n/3})$, less than classical attacks. And its data complexity is $O(2^{2n/3})$ with $O(2^{5n/6})$ classical memory and qubits.

Furthermore, we treat the above attack on 7-round as a new distinguisher. Similar to Dong \textit{et al.}'s attack \cite{Dong2018A}, we use Grover's algorithm to search the last $r-7$ rounds of subkeys as an outer loop with running time $2^{(r-7)n/4}$, and the above attack on the first 7 rounds as an inner loop with time $2^{2n/3}$. In total, the time complexity is only $O({2^{(r - 7)n/4 + 2n/3}})$ less than classical attacks and quantum attacks. Besides, our attack belongs to Q1 model, which is more practical. The detailed comparison with other attacks is shown in Table \ref{tab_compare_quantum_attack}.

\begin{table}[htbp]
	\centering
	\caption{Comparision with classical and quantum attacks on Feistel construction}
	\label{tab_compare_quantum_attack}
	\begin{tabular}{ccccccc}
		\hline
		Ref. &Setting &Round & Time &Data &Classcial memory & Qubits \\
		\hline
		\cite{Guo2016} &  Classical &6 & ${O(2^{3n/4})}$ & ${O(2^{3n/4})}$ &${O(2^{3n/4})}$ &- \\
		\cite{Zhao2019}  &  Classical &7 & ${O(2^n)}$ & ${O(2^n)}$ &${O(2^{3n/4})}$  &-\\
		\cite{Dong2020A}& Q2 & 4 & $O({n})$ &$O(2^{n/2})$ &- & $O(n)$  \\
		\cite{Bonnetain2020} & Q2 & 4 & $O({n^2})$ &$O(2^{n/2})$ &- & $O(n)$ \\
		\cite{Dong2018A}& Q2 & $r$ ($r \ge 3$) & $O({n2^{0.25nr - 0.75n}})$ &$O(2^{n/2})$&- & $O(n^2)$ \\
		\cite{Ito2019}& Q2 & $r$ ($r \ge 4$) & $O({n2^{0.25nr - n}})$ &$O(2^{n/2})$ &- & $O(n^2)$   \\
		\cite{Xie2019} & Q2 & 3 & $O({n^2})$ &$O(2^{n/2})$ &- & $O(n)$ \\
		\cite{Hosoyamada2018A} & Q1 & 6 & $O({2^{n/2}})$ &$O(2^{n/2})$ & - & $O({2^{n/2}})$  \\
		Ours & Q1 & $r$ ($r \ge 7$) & $O({2^{2n/3 + (r - 7)n/4}})$ & $O(2^{2n/3})$ & $O(2^{5n/6})$ & $O(2^{5n/6})$ \\
		\hline
	\end{tabular}
\end{table}

This paper is organized as follows. Sect. \ref{sec_related_alg} provides a brief description of related quantum algorithms. And an overview of Guo \textit{et al.}'s work is presented in Sect. \ref{sec_Classical_MITM}. We propose a new quantum meet-in-the-middle attack on 7-round Feistel construction in in Sect. \ref{sec_QMITM_7r}. Then, the attack is furthered on $r$-round ($r>7$) as shown in Sect. \ref{sec_QMITM_rr}, followed by a conclusion in Sect. \ref{sec_conclusion}.

\section{Related Quantum Algorithms}\label{sec_related_alg}

\noindent\textbf{Grover's Algorithm.} Grover's algorithm, or the Grover search, is one of the most famous quantum algorithms, with which we can obtain quadratic speed up on database searching problems compared to the classical algorithms. It was originally developed by Grover \cite{Grover1997}.

\begin{problem}\label{prob_grover}
	Suppose a function $f:{\{ 0,1\} ^n} \to \{ 0,1\} $ is given as a black box, with a promise that there is $x$ such that $f(x)=1$. Then, find $x$ such that $f(x)=1$.
\end{problem}

The process of solving above problem is presented simply as below by Grover's algorithm.
\begin{enumerate}
	\item Start with a uniform superposition $\left| \varphi  \right\rangle  = {1 \over {\sqrt {{2^n}} }}\sum\limits_{x \in {{\{ 0,1\} }^n}} {\left| x \right\rangle } $.
	
	\item The unitary transformation $\left( {2\left| \varphi  \right\rangle \left\langle \varphi  \right| - I} \right){\mathcal{O}_f}$ applied on $\left| \varphi  \right\rangle $ is iterated ${\sqrt {{2^n}} }$ times, where the quantum oracle ${\mathcal{O}_f}$ defined by the function $f$ is a quantum amplitude flip operation followed as bellow:
	\begin{equation}\label{eq_phase_flip}
		\mathcal{O}_f\left| x \right\rangle  \to\left\{
		\begin{aligned}
			- \left| x \right\rangle  & ,\  if \ f(x) = 1, \\
			\left| x \right\rangle  & ,\  otherwise.
		\end{aligned}
		\right.
	\end{equation}
	
	\item The final measurement gives the $x$ such that $f(x)=1$ with an overwhelming probability.
\end{enumerate}

In summary, Grover's algorithm can solve Problem \ref{prob_grover} with $O({2^{n/2}})$ evaluations of $f$ using $O(n)$ qubits. If we have $2^p$ independent small quantum processors with $O(n)$ qubits, by parallel running $O(\sqrt {{{{2^n}} \over {{2^p}}}} )$ iterations on each small quantum processor, we can find $x$ such that $f(x) = 1$ with high probability.  This parallelized algorithm runs in time $O(\sqrt {{{{2^n}} \over {{2^p}}}} )$.

\noindent\textbf{Amplitude Amplification.} Suppose there is a quantum algorithm $QSearch$ with the following properties. Let ${\mathcal{A}}$ be any quantum algorithm that uses no measurements, and boolean function $f:\mathbb{F}_2^n \to {\mathbb{F}_2}$. $p$ represents the success probability of ${\mathcal{A}}$ of finding a solution (i.e., the probability of outputting $x$ s.t. $f(x) = 1$). The algorithm $QSearch$ needs to execute $O(1/\sqrt p )$ times ${\mathcal{A}}$ and ${{\mathcal{A}}^{-1}}$ to find one solution if $p> 0$, otherwise it keeps running.

Generally speaking, $QSearch$ needs to iterate a certain number of unitary transformation $Q =  - {\mathcal{A}}{\mathcal{S}_0}{\mathcal{A}}{\mathcal{S}_f}$ on the initial state ${\mathcal{A}}\left| 0 \right\rangle $, where ${\mathcal{S}_{f}}\left| x \right\rangle  = {( - 1)^{f(x)}}\left| x \right\rangle $, ${\mathcal{S}_0}\left| 0 \right\rangle  =  - \left| 0 \right\rangle $ and ${\mathcal{S}_0}\left| x \right\rangle  = \left| x \right\rangle $ ($x \ne 0$). The algorithm $QSearch$ needs to iterate $O(1/\sqrt p )$ times $Q$ to get a solution with high probability a least $\max \{ p,1 - p\} $. We can obviously find that amplitude amplification can accelerate the search algorithm, and Grover's algorithm is a special case of amplitude amplification.

\noindent\textbf{Quantum Claw Finding Algorithm.} Quantum claw finding algorithm aims to solve the claw finding problem which is defined in detail as below.

\begin{problem}\label{prob_claw_finding}
	Suppose there are 2 functions $f:X \to Z$ and $g:Y \to Z$, where $\left| X \right| = M$ and $\left| Y \right| = N$. If there is a pair $(x,y) \in {X} \times {Y}$ such that $f(x) = g(y)$, then the pair is called a claw of the functions $f$ and $g$.
\end{problem}

To solve above problem, Buhrman et al.\cite{Buhrman2005} gave the first quantum claw finding algorithm based on amplitude amplification, which is shown in Algorithm \ref{algorithm_claw_finding}. 
\begin{algorithm}[t]
	\caption{Quantum Claw Finding between Two Arbitrary Functions $f$ and $g$ \cite{Buhrman2005}}
	\label{algorithm_claw_finding}
	{\bf Input:} Sets $X$ and $Y$ of size $N$ and $M$, respectively.
	
	{\bf Output:}  A pair of $(x,y) \in {X} \times {Y}$ such that $f(x) = g(y)$ if they exist.
	\begin{enumerate}
		\item Select a random subset $A \subseteq [N]$ of size $l$ ($l \le \min \{ N,\sqrt M \} $).
		
		\item Select a random subset $B \subseteq [M]$ of size $l^2$.
		
		\item Sort the elements in $A$ according to their $f$-value.	
		
		\item For a specific $b \in B$, check if there is an $a \in A$ such that $(a,b)$ is a claw by using classical binary search on the sorted version of $A$. Combine this with quantum search on the elements of $B$ to find a claw in $A \times B$.
		
		\item Apply amplitude amplification on steps 1-4.
	\end{enumerate}
\end{algorithm}

In Algorithm \ref{algorithm_claw_finding}, Step 3 requires classical sorting with $O({2^l} \cdot l)$ comparisons and Step 4 needs $O(\sqrt {\left| B \right|} \log \left| A \right|) = O(l\log l)$ comparisons, since checking if there is an $A$-element colliding with a given $b \in B$ takes $O(\log \left| A \right|) = O(\log l)$ comparisons via binary search on the sorted $A$, and quantum search needs $O(\sqrt {\left| B \right|} ) = O(l)$ to find a $B$-element that collides with an element occurring in $A$. In general, the time complexity of steps 1-4 is $O(l\log l)$. The probability of getting one claw $(x,y) \in A \times B$ in the first four steps is $p = {l^3}/2NM$. Therefore, the amplitude amplification of Step 5 requires an expected $O(\sqrt {NM/{l^3}} )$ iterations of steps 1-4. In total, the time complexity of steps 1-5 is $O(\sqrt {{{NM} \over l}} \log l)$. When $N \le M \le {N^2}$, $O(\sqrt {{{NM} \over l}} \log l) \approx O({N^{1/2}}{M^{1/4}}\log N)$. And when $M > {N^2}$, $O(\sqrt {{{NM} \over l}} \log l) \approx O({M^{1/2}}\log N)$.

\noindent\textbf{Quantum Random Access Memory.} A quantum random access memory (QRAM) is a quantum analogue of a classical random access memory (RAM), which uses $n$-qubit to address any quantum superposition of $2^n$ (quantum or classical) memory cells \cite{Giovannetti2008}. The QRAM is modeled as an unitary transformation ${U_{QRAM}}$ such that 
\begin{equation}\label{eq_qram}
	\sum\limits_i {{a_i}{{\left| i \right\rangle }_{addr}}} \buildrel {{U_{QRAM}}} \over
	\longrightarrow \sum\limits_i {{a_i}{{\left| i \right\rangle }_{addr}}{{\left| {{D_i}} \right\rangle }_{data}}},
\end{equation}
where $\sum\limits_i {{a_i}{{\left| i \right\rangle }_{addr}}} $ is a superposition of addresses and ${{D_i}}$ is the content of the $i$th memory cell.

\section{Overview of Classical Meet-in-the-Middle Attack on 6-Round Feistel Constructions}\label{sec_Classical_MITM}
Our proposed quantum meet-in-the-middle attack on Feistel constructions is based on Guo's work \cite{Guo2016}, so we first briefly introduce the framework of Guo's attack.

\subsection{Attack Idea} \label{Idea_CMITM}
\begin{figure}[htbp]
	\centering
	\includegraphics[width=8cm]{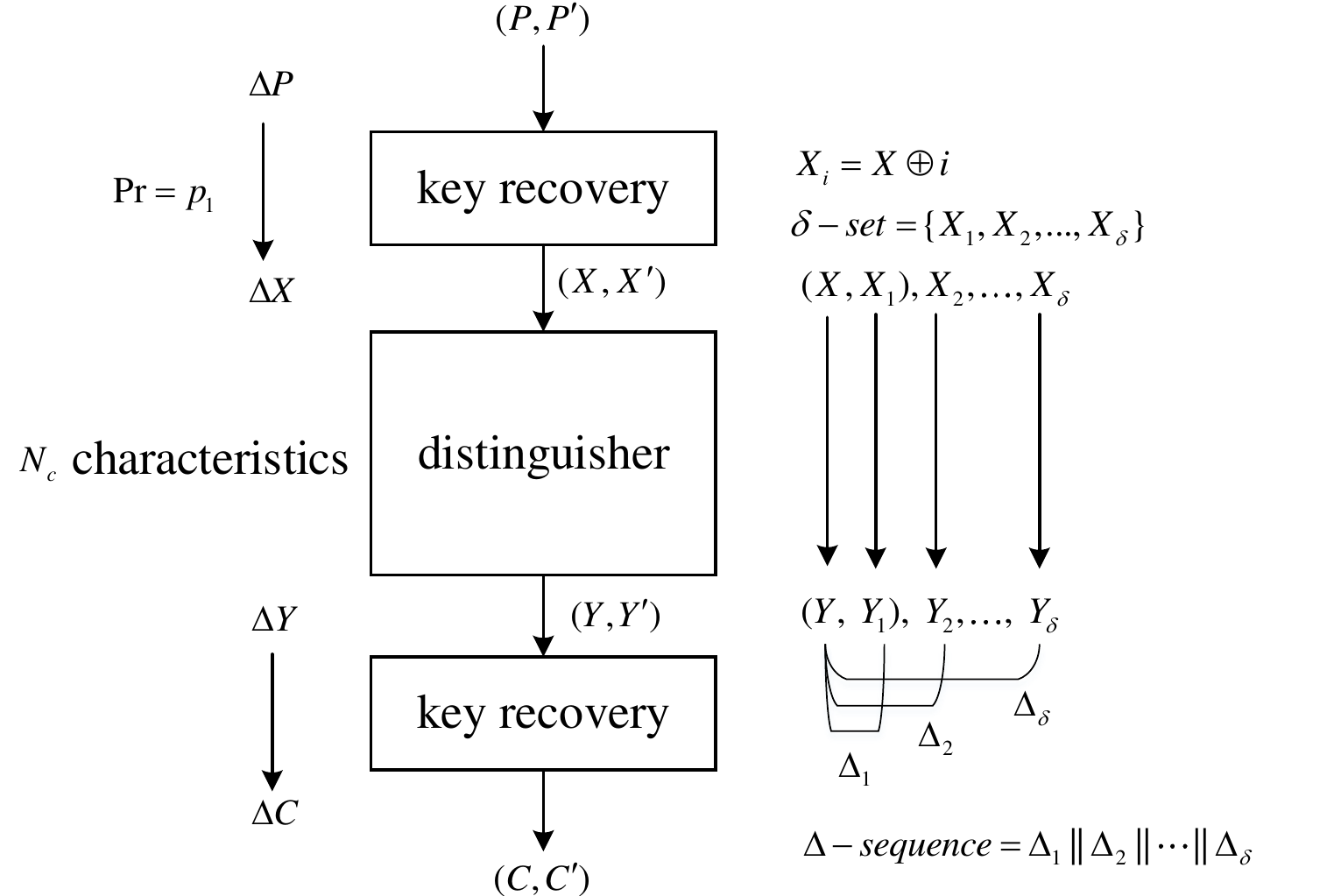}
	\caption{Overview of the MITM attack.}
	\label{fig_attack_idea}
\end{figure}

The MITM attack generally consists of the distinguisher and the key-recovery parts as illustrated in Fig. \ref{fig_attack_idea}. Suppose that a truncated differential is specified to the entire cipher and the plaintext difference $\Delta P$ propagates to the input difference $\Delta X$ of the distinguisher with probability ${p_1}$. Similarly, from the other direction,  the ciphertext difference $\Delta C$ propagates to the output difference $\Delta Y$ of the distinguisher with probability ${p_2}$. Generally, the attack consists of two phases: precomputation and queried-data analysis. 

In the pre-computation phase, the adversary firstly enumerates all the possible differential characteristics that can satisfy the truncated differential of the distinguisher. Suppose that there exit $N_c$ such characteristics. Let $(X,X')$ be the input pair values for each characteristic with difference $\Delta X$. Then, the adversary generates a $\delta$-set that contains ${X_1},{X_2}, \ldots ,{X_{\delta}}$, where ${X_i} = X \oplus i$ ($i = 1,2, \ldots ,\delta$). Let ${Y_1},{Y_2}, \ldots ,{Y_{\delta}}$ be the corresponding values at the output of the distinguisher. And the differences ${\Delta _i}$ between $Y$ and $Y_i$ for $i = 1,2, \ldots ,\delta$ make up a sequence, called $\Delta$-$sequence$ ($\Delta$-$sequence ={\Delta _1}||{\Delta _2}|| \cdots ||{\Delta _{\delta}}$). Note that the size of the difference ${\Delta _i}$ is $\lambda $ bits. In the end, $N_c$  $\Delta$-$sequence$s of the size $\lambda \delta $ bits would be stored in the Table ${T_\delta }$.

In the queried-data analysis phase, the adversary collects ${({p_1}{p_2})^{ - 1}}$ pairs of plaintexts with the difference $\Delta P$ and their corresponding ciphertexts with the difference $\Delta C$. One pair of collected pairs, with high probability, satisfies $\Delta X$ and $\Delta Y$ at the input and output of the distinguisher, respectively. Thus, for each of ${({p_1}{p_2})^{ - 1}}$ paired values, the adversary guesses subkeys for the key recovery rounds such that $\Delta X$ and $\Delta Y$ appear after the first and the last key recovery parts, respectively. Then, for each pair and the guessed subkeys, $P$ is modified to $P_0$ and the other ${P_1},{P_2}, \ldots ,{P_{\delta}}$ are computed by generated $\delta$-$set$ and the guessed subkeys. And these plaintexts are queried to obtain their corresponding ciphertexts. Next, the adversary partially decrypt these ciphertexts with the guessed subkeys, and the $\Delta$-$sequence$ is computed at the output of the distinguisher. Finally, those $\Delta$-$sequence$s are matched the table ${T_\delta }$, if the analyzed pair is a right pair and the guessed subkeys are correct, then a match will be found. Otherwise, a match will not be found as long as ${({p_1}{p_2})^{ - 1}}{N_c} \times {2^{ - \lambda \delta }} \ll 1$.

\subsection{Application on 6-Round Feistel Construction}\label{6r_CMITM}

\begin{figure}[htbp]
	\centering
	\includegraphics[width=12cm]{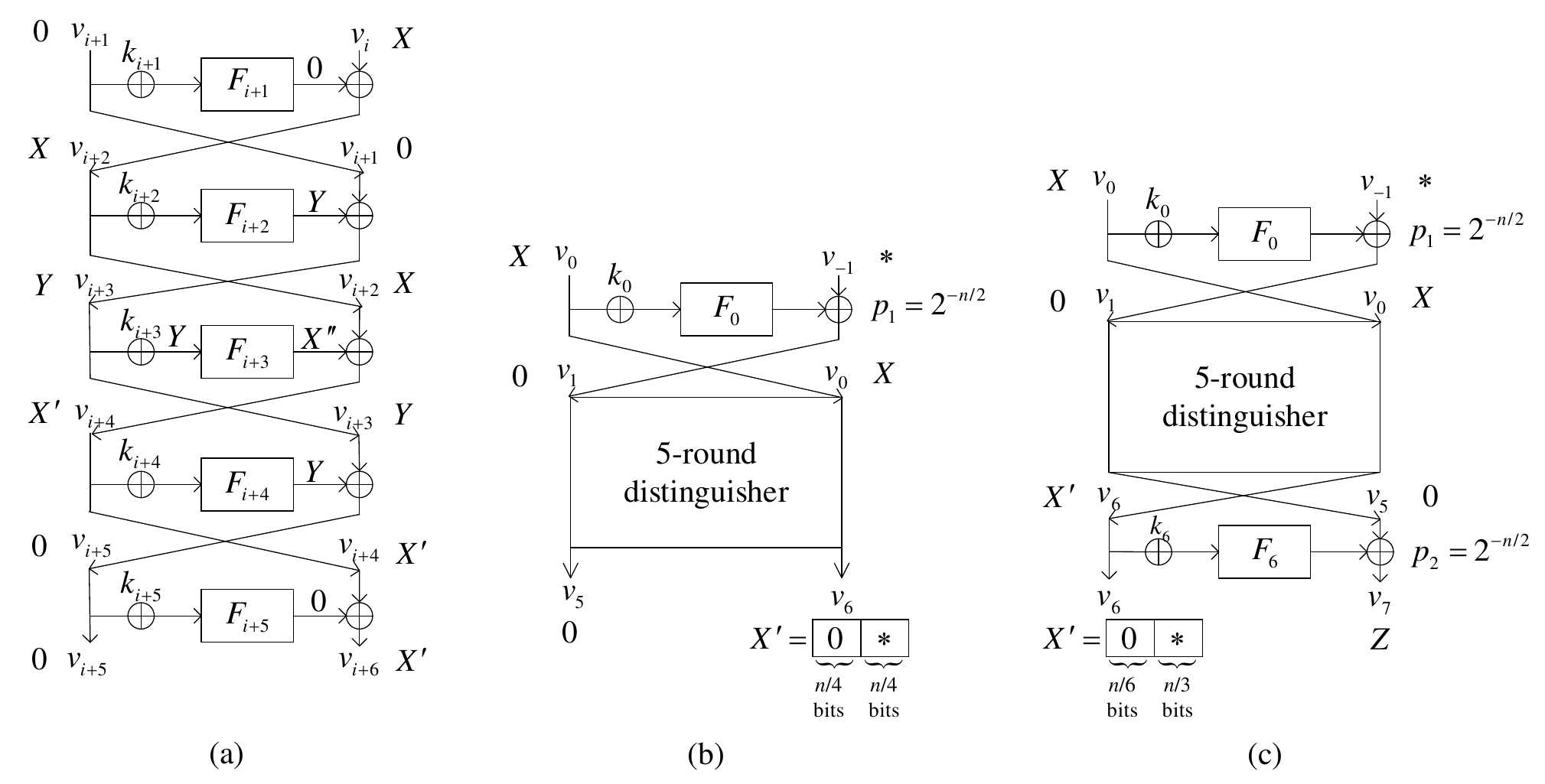}
	\caption{(a) The 5-round distinguisher. (b) 6-round Feistel construction for the key recovery. (c) 7-round Feistel construction for the key recovery. $v_i$ and $F_i$ represent the input value and the round-function of each round, respectively. $X'' = X' \oplus X$ and $X' \ne X$. $Z$ is the difference at ${v_7}$.}
	\label{fig_5r_distinguisher}
\end{figure}

\noindent\textbf{Pre-computation phase.} Guo \textit{et al.} \cite{Guo2016} give the 5-round distinguisher, which is illustrated in Fig. \ref{fig_5r_distinguisher}(a). The input and output differences of the 5-round distinguisher are defined as $0||X$ and $X'||0$, where $X,X' \in {\{ 0,1\} ^{n/2}}$, $ X \ne X'$ and the block size is $n$. For a given $X$, $X'$, the 5-round differential characteristics can be fixed to 
\begin{equation}\nonumber
	(0|| X)\buildrel {1stR} \over
	\longrightarrow ( X||0)\buildrel {2ndR} \over
	\longrightarrow ( Y|| X)\buildrel {3rdR} \over
	\longrightarrow ( X'|| Y)\buildrel {4thR} \over
	\longrightarrow (0||X')\buildrel {5thR} \over
	\longrightarrow (0||X'),
\end{equation}
where $Y$ represents the output difference of the 2nd round-function $F_{i+2}$ and has $2^{n/2}$ possible values. And the output difference of the 3rd round-function $F_{i+3}$ is $X''$($X'' = X \oplus X'$).

For one choice of $X$, $X'$, the number of the 5-round differential characteristics satisfying such input and output differences is $2^{n/2}$, i.e., the number of corresponding $\Delta$-$sequence$ for the left-half of the distinguisher's output is $2^{n/2}$ proved by the Proposition 1 in Ref. \cite{Guo2016}. To compute $\Delta$-$sequence$, we need to know the input values of the middle 3 rounds round-functions $F_{i+2}$,$F_{i+3}$, and $F_{i+4}$. According to the proof of the Lemma 1 in Ref. \cite{Guo2016}, for each $Y$, both input and output differences of $F_{i+2}$,$F_{i+3}$, and $F_{i+4}$ are fixed, which suggests that the paired values during the round-function are fixed to one choice on average. Thus, the adversary needs to construct three tables that record the input values of $F_{i+2}$,$F_{i+3}$, and $F_{i+4}$, for each $Y$ and one given $X$, $X'$, respectively. Finally, the computed $\Delta$-$sequence$s are stored in Table $T_\delta$ for for each $Y$ and one given $X$, $X'$, whose complexity is $2^{n/2}$ in both time and memory.

To balance the complexities between the pre-computation phase and the queried-data analysis phase, Guo \textit{et al.} iterate the above analysis for $2^{n/4}$ different choices of $X'$. They assume that the values of $X'$ differ in the last $n/4$ bits and are the same in the remaining $n/4$ bits.  Hence, the entire complexity of the pre-computation phase is $O(2^{3n/4})$ in both time and memory.

\noindent\textbf{Queried-data analysis phase.} Guo \textit{et al.} append 1-round before the 5-round distinguisher to attack the 6-round Feistel construction, as illustrated in the Fig. \ref{fig_5r_distinguisher}(b). Firstly, the adversary prepares 2 sets of plaintexts in the form of $\{ (m||0),(m||1), \ldots ,(m||{2^{n/2 - 1}})\} $ and $\{ (m \oplus X||0),(m \oplus X||1), \ldots ,(m \oplus X||{2^{n/2 - 1}})\} $ respectively, where $m$ is a randomly chosen $n/2$-bit constant. These plaintexts can compose $2^n$ pairs, and only $2^{n/4}$ pairs satisfy $\Delta C$ in the corresponding ciphertexts for $2^{n/4}$ choices of $X'$. By iterating this procedure $2^{n/4}$ times for different choice of $m$, $2^{n/2}$ pairs satisfying $\Delta C$ are collected. Due to that the probability $p_1$ that a randomly chosen plaintext pair with the difference $X|| * $ satisfies the difference $0||X$ after 1 round is $2^{-n/2}$, one pair will satisfy the entire differential characteristic. $* $ can be any $n/2$-bit difference.

Since we assume that the input difference of the distinguisher is $0||X$ for each pair, the input and output differences of $F_0$ are fixed, $X$ and $* $, which will fix one key candidate for $k_0$ uniquely. The adversary computes a set of plaintexts by the $\delta$-set and the key candidate for each pair, and queries their corresponding ciphertexts. Finally, The adversary computes the corresponding $\Delta$-$sequence$ and matches $T_\delta$. If there is a match, the key candidate for $k_0$ is correct and the $Y$ can be known. The other keys are trivially recovered from the second round one by one. Note that ${p_1} = {2^{ - n/2}}$, ${p_2} = 1$, $N_c=2^{n/2}$, $\lambda  = n/2$, hence, $\delta  = 3$ is sufficient to filter out all the wrong candidates.

\section{Quantum Meet-in-the-Middle Attack on 7-round Feistel Construction}\label{sec_QMITM_7r}
To attack 7-round Feistel, we append 1-round before and after the 5-round distinguisher respectively, and the differential characteristic of the 7-round Feistel that we need is illustrated in the Fig. \ref{fig_5r_distinguisher} (c). Similar to classic meet-in-the-middle attack, we also divide our attack into two phase: pre-computation phase and queried-data analysis phase.

\subsection{Pre-computation Phase}\label{pre-computation}
Similar to the classical MITM on 6-round Feistel, we need to calculate the $\Delta$-$sequence$ based on the distinguisher and choose a fixed $X$ value. But, $2^{n/3}$ choices of $X'$ are considered, to balance the complexities between the pre-computation phase and the queried-data analysis phase. Without loss of generality, assume that the values of $X'$ differ in the last $n/3$ bits and are zero in the remaining $n/6$ most significant bits (MSBs). We define a function $\mathcal{F}:X' \times Y \to \Delta $-$sequence$, where $X',Y \in {\{ 0,1\} ^{n/2}}$, $0 \le X' \le {2^{n/3}} - 1$, $0 \le Y \le {2^{n/2}} - 1$ and $\Delta$-$sequence$ $\in {\{ 0,1\} ^{\delta n/2}}$.

For a given $X'$ and $Y$, assume that a pair of plaintexts $(m,m \oplus 0||X)$ satisfy the differential characteristic of the distinguisher, ${t_{i + 2}}$, ${t_{i + 3}}$ and ${t_{i + 4}}$ are the input values of $F_{i+2}$, $F_{i+3}$ and $F_{i+4}$. Let us consider a new pair of plaintexts $(m,m \oplus 0||j )$($j \in \{1,2, \ldots ,\delta\} $) and compute the corresponding ${\Delta _j}$ that is the difference at $v_{i+5}$. 

Since, the output difference $\Delta {F_{i + 1}^\mathcal{O}}$ of the first round-function ${F_{i + 1}}$ in the distinguisher is always 0. So, $\Delta {v_{i+2}} = \Delta {v_i} = j$. In the second round, due to the input value $F_{i+2}^\mathcal{I} = {t_{i+2}}$ and $\Delta F_{i+2}^\mathcal{I} = j$, then, $\Delta F_{i+2}^\mathcal{O} = {F_{i+2}}({t_{i+2}}) \oplus {F_{i+2}}({t_{i+2}} \oplus j)$. And in the 3rd round, $\Delta F_{i+3}^\mathcal{O} = {F_{i+3}}({t_{i+3}}) \oplus {F_{i+3}}({t_{i+3}} \oplus \Delta F_{i+2}^\mathcal{O})$ because of $F_{i+3}^\mathcal{I} = {t_{i+3}}$ and $\Delta F_{i+3}^\mathcal{I} = \Delta F_{i+2}^\mathcal{O}$. In the next round, $\Delta {v_{i+4}} = \Delta F_{i+3}^\mathcal{O} \oplus j$, the output difference of 4th round is $\Delta F_{i+4}^\mathcal{O} = {F_{i+4}}({t_{i+4}}) \oplus {F_{i+4}}({t_{i+4}} \oplus \Delta F_{i+3}^\mathcal{O}\oplus j)$. Finally, we can get the difference ${\Delta _j}$ at ${v_5}$, 
\begin{equation}\label{eq_delta_j}
	\begin{split}
		{\Delta _j} &= \Delta {v_{i + 5}^j}\\
		&=\Delta F_{i + 4}^\mathcal{O} \oplus \Delta F_{i + 2}^\mathcal{O}\\
		&={F_{i + 4}}({t_{i + 4}}) \oplus {F_{i + 4}}({t_{i + 4}} \oplus \Delta F_{i + 3}^{\cal O} \oplus j) \oplus \Delta F_{i + 2}^\mathcal{O}\\
		&={F_{i + 4}}({t_{i + 4}}) \oplus {F_{i + 4}}({t_{i + 4}} \oplus \Delta F_{i + 3}^{\cal O} \oplus j) \oplus {F_{i + 2}}({t_{i + 2}}) \oplus {F_{i + 2}}({t_{i + 2}} \oplus j)\\
		&={F_{i + 4}}({t_{i + 4}}) \oplus {F_{i + 4}}({t_{i + 4}} \oplus {F_{i + 3}}({t_{i + 3}}) \oplus {F_{i + 3}}({t_{i + 3}} \oplus {F_{i + 2}}({t_{i + 2}})\\
		& \oplus {F_{i + 2}}({t_{i + 2}} \oplus j)) \oplus j) \oplus {F_{i + 2}}({t_{i + 2}}) \oplus {F_{i + 2}}({t_{i + 2}} \oplus j).
	\end{split}
\end{equation}
By repeating this procedure for different choices of $j$, we can get 
\begin{equation}\label{eq_delta_sequence}
	\Delta-sequence=\Delta_1||\Delta_2||...||\Delta_{\delta}
\end{equation}
for a given $X'$ and $Y$. 

From Eq. \ref{eq_delta_j}, to calculate ${\Delta _j}$, we need to know the values of ${t_{i + 2}}$, ${t_{i + 3}}$ and ${t_{i + 4}}$ corresponding to the given ($X$, $X'$, $Y$). According to the Lemma 1 in Ref. \cite{Guo2016}, there exists one state value (one solution) that satisfies such input-output difference in each of the middle three rounds for a given ($X$, $X'$, $Y$). As $Y$ takes at most $2^{n/2}$ different values, ${t_{i + 2}}$, ${t_{i + 3}}$ and ${t_{i + 4}}$ can assume only $2^{n/2}$ different values. Therefore, we need to search for eligible ${t_{i + 2}}$, ${t_{i + 3}}$ and ${t_{i + 4}}$ from a space of size $2^{n/2}$ for a given ($X$, $X'$, $Y$), where ${t_{i + 2}}$, ${t_{i + 3}}$ and ${t_{i + 4}}$ satisfy 
\begin{equation}\label{eq_t_i}
	\begin{split}
		&{F_{i + 2}}({t_{i + 2}}) \oplus {F_{i + 2}}({t_{i + 2}} \oplus X) = Y,\\
		&{F_{i + 3}}({t_{i + 3}}) \oplus {F_{i + 3}}({t_{i + 3}} \oplus Y) = X \oplus X',\\
		&{F_{i + 4}}({t_{i + 4}}) \oplus {F_{i + 4}}({t_{i + 4}} \oplus X') = Y.
	\end{split}
\end{equation}

So far, the above process is the calculation process of the function $\mathcal{F}$. In the classical setting, the final $\Delta$-$sequence$s are stored in a table $T_\delta$ with $O({2^n})$ computations and $O({2^n})$ classical memory. With quantum parallelism and superposition, we can reduce the time complexity to $O({2^{n/4}})$ with $O(n)$ qubits, which is shown as below.
\begin{enumerate}
	\item Prepare the superposition state
	\begin{equation}\label{eq_delta_initial}
		\left| {{\varphi _1}} \right\rangle  = \sum\limits_{X' = 0}^{{2^{n/3}} - 1} {{1 \over {\sqrt {{2^{n/3}}} }}{{\left| {X'} \right\rangle }_1}} \sum\limits_{Y = 0}^{{2^{n/2}} - 1} {{1 \over {\sqrt {{2^{n/2}}} }}{{\left| Y \right\rangle }_2}} {\left| X \right\rangle _3},
	\end{equation}
	where the bit length of $X'$ is $n/2$ and $X' \ne X$.
	
	\item Do Grover search for ${t_{i + 2}}$, ${t_{i + 3}}$ and ${t_{i + 4}}$ which satisfy Eq. \ref{eq_t_i} for each ($X'$, $Y$), and store them in the new registers, i.e.
	\begin{equation}\label{eq_grover}
		\begin{split}
			&\sum\limits_{Y = 0}^{{2^{n/2}} - 1} {{1 \over {\sqrt {{2^{n/2}}} }}{{\left| Y \right\rangle }_2}} {\left| X \right\rangle _3}\sum\limits_{{t_{i + 2}} = 0}^{{2^{n/2}} - 1} {{{\left| {{t_{i + 2}}} \right\rangle }_4}} \buildrel {Grover} \over
			\longrightarrow \sum\limits_{Y = 0}^{{2^{n/2}} - 1} {{1 \over {\sqrt {{2^{n/2}}} }}{{\left| Y \right\rangle }_2}} {\left| X \right\rangle _3}{\left| {{t_{i + 2}}} \right\rangle _4},\\
			&\sum\limits_{X' = 0}^{{2^{n/3}} - 1} {{1 \over {\sqrt {{2^{n/3}}} }}{{\left| {X'} \right\rangle }_1}} \sum\limits_{Y = 0}^{{2^{n/2}} - 1} {{1 \over {\sqrt {{2^{n/2}}} }}{{\left| Y \right\rangle }_2}} {\left| X \right\rangle _3}\sum\limits_{{t_{i + 3}} = 0}^{{2^{n/2}} - 1} {{{\left| {{t_{i + 3}}} \right\rangle }_5}} \buildrel {Grover} \over\longrightarrow \\
			&\sum\limits_{X' = 0}^{{2^{n/3}} - 1} {{1 \over {\sqrt {{2^{n/3}}} }}{{\left| {X'} \right\rangle }_1}} \sum\limits_{Y = 0}^{{2^{n/2}} - 1} {{1 \over {\sqrt {{2^{n/2}}} }}{{\left| Y \right\rangle }_2}} {\left| X \right\rangle _3}{\left| {{t_{i + 3}}} \right\rangle _5}\\
			&\sum\limits_{X' = 0}^{{2^{n/3}} - 1} {{1 \over {\sqrt {{2^{n/3}}} }}{{\left| {X'} \right\rangle }_1}} \sum\limits_{Y = 0}^{{2^{n/2}} - 1} {{1 \over {\sqrt {{2^{n/2}}} }}{{\left| Y \right\rangle }_2}} \sum\limits_{{t_{i + 4}} = 0}^{{2^{n/2}} - 1} {{{\left| {{t_{i + 4}}} \right\rangle }_6}} \buildrel {Grover} \over\longrightarrow\\
			& \sum\limits_{X' = 0}^{{2^{n/3}} - 1} {{1 \over {\sqrt {{2^{n/3}}} }}{{\left| {X'} \right\rangle }_1}} \sum\limits_{Y = 0}^{{2^{n/2}} - 1} {{1 \over {\sqrt {{2^{n/2}}} }}{{\left| Y \right\rangle }_2}} {\left| {{t_{i + 4}}} \right\rangle _6}.
		\end{split}
	\end{equation} 
	And we obtain
	\begin{equation}\label{eq_grover_final}
		\left| {{\varphi _2}} \right\rangle  = \sum\limits_{X',Y = 0}^{X'= {2^{n/3}} - 1,Y = {2^{n/2}} - 1} {{2^{ - 5n/12}}{{\left| {X'} \right\rangle }_1}{{\left| Y \right\rangle }_2}} {\left| X \right\rangle _3}{\left| {{t_{i + 2}}} \right\rangle _4}{\left| {{t_{i + 3}}} \right\rangle _5}{\left| {{t_{i + 4}}} \right\rangle _6}
	\end{equation} 
	after Grover search with time $O(2^{n/4})$.
	
	\item Compute $\Delta$-$sequence$ according to Eq. \ref{eq_delta_j} with $\mathop  \otimes \limits_{j = 1}^\delta  \left| j \right\rangle $ as input, and store them in a new register of $\delta n/2$ qubits.
	\begin{equation}\label{eq_q_delta}
		\begin{split}
			\left| {{\varphi _3}} \right\rangle  &= \sum\limits_{X',Y = 0}^{X'= {2^{n/3}} - 1,Y = {2^{n/2}} - 1} {{2^{ - 5n/12}}{{\left| {X'} \right\rangle }_1}{{\left| Y \right\rangle }_2}} {\left| X \right\rangle _3}{\left| {\Delta  - sequence} \right\rangle _7}\\
			&= \sum\limits_{X',Y = 0}^{X'= {2^{n/3}} - 1,Y = {2^{n/2}} - 1} {{2^{ - 5n/12}}{{\left| {X'} \right\rangle }_1}{{\left| Y \right\rangle }_2}} {\left| X \right\rangle _3}{\left| {{\cal F}(X',Y)} \right\rangle _7},
		\end{split}
	\end{equation}
	where we discard some useless registers finally.
	
	\item Do Grover search parallelly on $\left| {{\varphi _3}} \right\rangle$ for each $(X',Y)$ with $2^{5n/6}$ parallel single quantum processors, and store the measured $\left| {\Delta  - sequence} \right\rangle _7$ in the table $T_\delta$ indexed by $(X',Y)$.
\end{enumerate}

In summary, the entire quantum computation for the table $T_\delta$ will cost $O(2^{2n/3})$ time with $O(2^{5n/6})$ qubits and $O(2^{5n/6})$ classical memory.

\subsection{Queried-data analysis phase}\label{query_phase}
Firstly, we choose two plaintext sets in the form of $\{ m||0,m||1,...,m||{2^{n/2}} - 1\} $ and $\{ m \oplus X||0,m \oplus X||1,...,m \oplus X||{2^{n/2}} - 1\} $, where $m$  is a randomly chosen $n/2$-bit constant. ${2^n}$ pairs of plaintexts can be generated, $2^{5n/6}$ pairs of them will satisfy $\Delta C = X'||Z$ in the corresponding ciphertexts for $2^{n/3}$ choices of $X'$ and $2^{n/2}$ choices of $Z$. By iterating this procedure $2^{n/6}$ times for different choices of $m$, $2^n$ pairs satisfying $\Delta P = X|| * $ and $\Delta C = X'||Z$ are collected. These pairs are stored in the table $T_{PC}$ indexed by the item $I$ ($0 \le I \le {2^n} - 1$). We can see that, the probability of that the plaintext difference $\Delta P$ propagates to the input difference $0||X$ of the distinguisher is ${p_1} = {2^{ - n/2}}$, and the probability of the ciphertext difference $\Delta C$ propagates to the output difference $X'||0$ is ${p_2} = {2^{ - n/2}}$. Therefore, one pair of $2^n$ pairs would satisfy the entire differential characteristic of 7-round Feistel.

In this phase, we aim to compute the $\Delta$-$sequence$ for each pair, and we define this computing process as a function ${\cal G}:I \to$$\Delta$-$sequence$, where $0 \le I \le {2^n} - 1$ and $\Delta$-$sequence$$ \in {\{ 0,1\} ^{\delta n/2}}$.

For each pair, we should guess subkeys $k_0$ and $k_6$ to ensure that the differences $\Delta P$ and $\Delta C$ propagate to the input difference $0||X$ and the output difference $X'||0$ of the distinguisher respectively. In other words, we assume that the differences $\Delta P$ and $\Delta C$ for each pair propagate to $0||X$ and $X'||0$ respectively. In this case, we can find the input values of the two round-functions $F_0$ and $F_6$ according to the known input and output differences, and then use the input values with the known plaintext and ciphertext values to deduce the guessed subkeys, called subkey candidates.

Suppose that one pair of $2^n$ pairs is $(P,P')=({v_0}||{v_{ - 1}},{v_0} \oplus X||{v_{ - 1}} \oplus  * )$ and their corresponding ciphertexts $(C,C')=({v_6}||{v_7},{v_6} \oplus X'||{v_7} \oplus Z)$. Then, we need to search the input values $F_0^\mathcal{I}$ and $F_6^\mathcal{I}$ according to the input and output differences. Therefore, we compute all $2^{n/2}$ possible values of $F_0^\mathcal{I}$ and $F_6^\mathcal{I}$ by Eq. \ref{eq_F0_F6}, and store them in two tables $T_0$ and $T_6$ indexed by $(X,*)$ and $(X',Z)$ respectively. This step will cost $O(2^{5n/6})$ time classically.
\begin{equation}\label{eq_F0_F6}
	{F_0}(F_0^\mathcal{I}) \oplus {F_0}(F_0^\mathcal{I} \oplus X) =  *, \  {F_6}(F_6^\mathcal{I}) \oplus {F_6}(F_6^\mathcal{I} \oplus X') = Z .
\end{equation}
When we get $F_0^\mathcal{I}$ and $F_6^\mathcal{I}$, we can obtain two subkeys candidates $k_0$ and $k_6$ by
\begin{equation}\label{eq_k0_k6}
	{k_0} = F_0^\mathcal{I} \oplus {v_0}, \  {k_6} = F_6^\mathcal{I} \oplus {v_6} .
\end{equation}

Then, we construct a $\delta$-set $\{ {v_1}||{v_0} \oplus 1,{v_1}||{v_0} \oplus 2,...,{v_1}||{v_0} \oplus \delta \} $, where ${v_1}||{v_0}$ is corresponding to the plaintext ${v_0}||{v_{-1}}$ and can be obtained by encrypting with the subkey candidate $k_0$. So, with the knowledge of subkey candidate $k_0$, we compute the corresponding $\Delta F^\mathcal{O}_0$, modify $v_{-1}$ so that the value of $v_1$ stays unchanged, and obtain the plaintexts corresponding to the $\delta$-set, i.e.,
\begin{equation}\label{eq_P_j}
	\begin{split}
		\Delta F_0^{\cal O}& = {F_0}({v_0} \oplus {k_0}) \oplus {F_0}({v_0} \oplus {k_0} \oplus j), \ j \in \{ 1,2,...,\delta \}, \\
		{P_j}& = {v_0} \oplus j||{v_{ - 1}} \oplus \Delta F_0^{\cal O}.\\
	\end{split}
\end{equation}
And these plaintexts are queried to get ciphertexts $\{ {C_1},{C_2},...,{C_\delta }\} $. With the knowledge of subkey candidate $k_6$, these ciphertexts are decrypted partially to get the values $\{ v_5^1,v_5^2,...,v_5^\delta \} $ at $v_5$. Finally, the differences between $v_5$ (corresponding to ciphertext $C$) and $v_5^j$ make up a sequence $\Delta$-$sequence$. If this $\Delta$-$sequence$ can be found in Table $T_\delta$, the subkey candidates $k_0$ and $k_6$ are correct. Note that, ${p_1} = {2^{ - n/2}}$, ${p_2} = {2^{ - n/2}}$, $N_c=2^{n/2}$, $\lambda  = n/2$, hence, $\delta  = 4$ is sufficient to filter out all the wrong candidates.

In the classical setting, the computing process of $\mathcal{G}$ needs $O({2^n})$ computations with $O({2^n})$ classical memory. To reduce complexity, we do the following quantum operations.
\begin{enumerate}
	\item Prepare the superposition state
	\begin{equation}\label{eq_initial_I}
		\left| {{\varphi _1}} \right\rangle  = \sum\limits_{I = 0}^{{2^n} - 1} {{1 \over {\sqrt {{2^n}} }}{{\left| I \right\rangle }_1}} .
	\end{equation}
	
	\item Use $\left| {{\varphi _1}} \right\rangle$ as the address to query the paired plaintexts and ciphertexts $(P,P',C,C')$ from the table $T_{PC}$ by QRAM,
	\begin{equation}\label{eq_initial_I_PC}
		\begin{split}
			\left| {{\varphi _2}} \right\rangle  &= \sum\limits_{I = 0}^{{2^n} - 1} {{1 \over {\sqrt {{2^n}} }}{{\left| I \right\rangle }_1}} {\left| {P,P',C,C'} \right\rangle _2}\\
			&=\sum\limits_{I = 0}^{{2^n} - 1} {{1 \over {\sqrt {{2^n}} }}{{\left| I \right\rangle }_1}} {\left| {PC} \right\rangle _2},\\
		\end{split}
	\end{equation}
	where $P,P',C,C'$ is abbreviated as  $PC$.
	
	\item Compute $(X,*,X',Z)$ for each pair to obtain
	\begin{equation}\label{eq_initial_I_PC_differ}
		\left| {{\varphi _3}} \right\rangle  = \sum\limits_{I = 0}^{{2^n} - 1} {{1 \over {\sqrt {{2^n}} }}{{\left| I \right\rangle }_1}} {\left| {PC} \right\rangle _2}{\left| {X, * ,X',Z} \right\rangle _3}
	\end{equation}
	
	\item Do Grover search to find $F_0^\mathcal{I}$ and $F_6^\mathcal{I}$ satisfying Eq. \ref{eq_F0_F6} for each pair
	\begin{equation}\label{eq_initial_I_PC_F0_F6}
		\begin{split}
			&\sum\limits_{I = 0}^{{2^n} - 1} {{1 \over {\sqrt {{2^n}} }}{{\left| I \right\rangle }_1}} {\left| {PC} \right\rangle _2}{\left| {X, * ,X',Z} \right\rangle _3}\sum\limits_{F_0^\mathcal{I} = 0}^{{2^{n/2}} - 1} {{{\left| {F_0^\mathcal{I}} \right\rangle }_4}} \buildrel {Grover} \over
			\longrightarrow\\
			& \sum\limits_{I = 0}^{{2^n} - 1} {{1 \over {\sqrt {{2^n}} }}{{\left| I \right\rangle }_1}} {\left| {PC} \right\rangle _2}{\left| {X, * ,X',Z} \right\rangle _3}{\left| {F_0^\mathcal{I}} \right\rangle _4}\\
			&\sum\limits_{I = 0}^{{2^n} - 1} {{1 \over {\sqrt {{2^n}} }}{{\left| I \right\rangle }_1}} {\left| {PC} \right\rangle _2}{\left| {X, * ,X',Z} \right\rangle _3}\sum\limits_{F_6^\mathcal{I} = 0}^{{2^{n/2}} - 1} {{{\left| {F_6^\mathcal{I}} \right\rangle }_5}} \buildrel {Grover} \over
			\longrightarrow\\
			&\sum\limits_{I = 0}^{{2^n} - 1} {{1 \over {\sqrt {{2^n}} }}{{\left| I \right\rangle }_1}} {\left| {PC} \right\rangle _2}{\left| {X, * ,X',Z} \right\rangle _3}{\left| {F_6^\mathcal{I}} \right\rangle _5},
		\end{split}
	\end{equation}
	which costs $O(2^{n/4})$ time, and obtain 
	\begin{equation}\label{eq_initial_I_PC_F0_F6_Grover}
		\left| {{\varphi _4}} \right\rangle  = \sum\limits_{I = 0}^{{2^n} - 1} {{1 \over {\sqrt {{2^n}} }}{{\left| I \right\rangle }_1}} {\left| {PC} \right\rangle _2}{\left| {X, * ,X',Z} \right\rangle _3}{\left| {F_0^\mathcal{I}} \right\rangle _4}{\left| {F_6^\mathcal{I}} \right\rangle _5}.
	\end{equation}

	\item Compute the subkey candidates $k_0$ and $k_6$ with known plaintext $P$ and ciphertext $C$ according to Eq. \ref{eq_k0_k6}, to get 
	\begin{equation}\label{eq_initial_I_PC_k0_k6}
		\left| {{\varphi _5}} \right\rangle  = \sum\limits_{I = 0}^{{2^n} - 1} {{1 \over {\sqrt {{2^n}} }}{{\left| I \right\rangle }_1}} {\left| {PC} \right\rangle _2}{\left| {X, * ,X',Z} \right\rangle _3}{\left| {F_0^I} \right\rangle _4}{\left| {F_6^I} \right\rangle _5}{\left| {{k_0}} \right\rangle _6}{\left| {{k_6}} \right\rangle _7}.
	\end{equation}
	
	\item Compute the plaintexts corresponding to the $\delta$-set according to Eq. \ref{eq_P_j}, to get 
	\begin{equation}\label{eq_initial_I_PC_P_j}
		\left| {{\varphi _6}} \right\rangle  = \sum\limits_{I = 0}^{{2^n} - 1} {{1 \over {\sqrt {{2^n}} }}{{\left| I \right\rangle }_1}} {\left| {PC} \right\rangle _2}{\left| {X, * ,X',Z} \right\rangle _3}{\left| {F_0^I} \right\rangle _4}{\left| {F_6^I} \right\rangle _5}{\left| {{k_0}} \right\rangle _6}{\left| {{k_6}} \right\rangle _7}{\left| {{P_1},{P_2}, \ldots .{P_\delta }} \right\rangle _8}.
	\end{equation}
	
	\item Query the corresponding ciphertexts $\left\{ {{C_1},{C_2},...,{C_\delta }} \right\}$ for $\left\{ {{P_1},{P_2}, \ldots .{P_\delta }} \right\}$ by QRAM, 
	\begin{equation}\label{eq_initial_I_PC_C_j}
		\begin{split}
			\left| {{\varphi _7}} \right\rangle  = \sum\limits_{I = 0}^{{2^n} - 1} {{1 \over {\sqrt {{2^n}} }}}&{{\left| I \right\rangle }_1} {\left| {PC} \right\rangle _2}{\left| {X, * ,X',Z} \right\rangle _3}{\left| {F_0^I} \right\rangle _4}{\left| {F_6^I} \right\rangle _5}{\left| {{k_0}} \right\rangle _6}{\left| {{k_6}} \right\rangle _7}\\
			& \otimes {\left| {{P_1},{P_2}, \ldots .{P_\delta }} \right\rangle _8}{\left| {{C_1},{C_2}, \ldots ,{C_\delta }} \right\rangle _9}.\\
		\end{split}
	\end{equation}
	Note that another table $T'_{PC}$ stores $2^{2n/3+1+\delta}$ plaintexts and corresponding ciphertexts, which is indexed by the plaintexts.
	
	\item Compute the $\Delta$-$sequence$ with the knowledge of subkey candidate $k_6$ and ciphertexts $\left\{ {{C_1},{C_2},...,{C_\delta }} \right\}$, to obtain
	\begin{equation}\label{eq_initial_I_PC_Delta}
		\begin{split}
			\left| {{\varphi _8}} \right\rangle & = \sum\limits_{I = 0}^{{2^n} - 1} {{1 \over {\sqrt {{2^n}} }}{{\left| I \right\rangle }_1}} {\left| {\Delta  - sequence} \right\rangle _{10}}\\
			&=\sum\limits_{I = 0}^{{2^n} - 1} {{1 \over {\sqrt {{2^n}} }}{{\left| I \right\rangle }_1}} {\left| {\mathcal{G}(I)} \right\rangle _{10}},
		\end{split}
	\end{equation}
	where we discard some useless registers.
\end{enumerate}

For now, we need to find a claw $\left( {\left( {X',Y} \right),I} \right)$ such that $\mathcal{F}(X',Y)=\mathcal{G}(I)$ from the table $T_\delta$ and the superposition state $\sum\limits_{I = 0}^{{2^n} - 1} {{1 \over {\sqrt {{2^n}} }}{{\left| I \right\rangle }_1}} {\left| {\mathcal{G}(I)} \right\rangle _{10}}$, which can be implemented by Algorithm \ref{algorithm_claw_finding_F_G}. When we find the claw $\left( {\left( {X',Y} \right),I} \right)$, the corresponding subkeys candidates $k_0$ and $k_6$ are correct, and the other subkeys can be deduced with the knowledge of $X$, $X'$ and $Y$.
\begin{algorithm}[t]
	\caption{Quantum Claw Finding between Two Functions $\mathcal{F}$ and $\mathcal{G}$}
	\label{algorithm_claw_finding_F_G}
	\begin{enumerate}
		\item Select a random subset $A \subseteq \{ (X',Y)\} $ of size $l$ ($l \le {2^{5n/6}}$), where $\left| {\{ (X',Y)\} } \right| = {2^{5n/6}}$.
		
		\item Select a random subset $B \subseteq \{ I\}$ of size $l^2$.
		
		\item Compute $\sum\limits_{b \in B} {{1 \over l}\left| b \right\rangle } \left| {\mathcal{G}(b)} \right\rangle $ according to Eqs. \ref{eq_initial_I}-\ref{eq_initial_I_PC_Delta}.
		
		\item Sort the elements in $A$ according to their $\mathcal{F}$-value according to the table $T_\delta$.	
		
		\item For a specific $b \in B$, check if there is an $a \in A$ such that $(a,b)$ is a claw by using classical binary search on the sorted version of $A$. Combine this with quantum search on the elements of $B$ to find a claw in $A \times B$.
		
		\item Apply amplitude amplification on steps 1-4 with $\sqrt {{{{2^{5n/6 + n}}} \over {{l^3}}}} $ iterations. And output the claw $\left( {\left( {X',Y} \right),I} \right)$ such that $\mathcal{F}(X',Y)=\mathcal{G}(I)$.
	\end{enumerate}
\end{algorithm}

\noindent\textbf{Complexity.} Obviously, our attack belongs to Q1 model, which needs $O(2^{2n/3})$ classical queries. And the time complexities of constructing the table $T_\delta$ and Algorithm \ref{algorithm_claw_finding_F_G} are both $O(2^{2n/3})$. Therefore, the total time complexity of our attack is $O(2^{2n/3})$. Besides, our attack needs $O(2^n)$ classical memory due to the tables $T_\delta$, $T_{PC}$ and $T'_{PC}$. And it consumes $O(2^{5n/6})$ qubits.

\section{Quantum Meet-in-the-Middle Attack on $r$-round Feistel Construction}\label{sec_QMITM_rr}
To attack $r$-round ($r > 7$) Feistel construction, Dong \textit{et al.} \cite{Dong2018A} guess the subkeys of the last $r-3$ rounds by Grover search, and use Simon's algorithm to recover the period satisfying the first 3-round distinguisher, which is inspired by Leander and May's work \cite{Leander2017}. Furthermore, Ito \textit{et al.} \cite{Ito2019} continue this attack idea and propose a new 4-round distinguisher to reduce the time complexity. Similarly, we guess the subkeys of the last $r-7$ rounds by Grover search, and find a claw between functions $\mathcal{F}$ and $\mathcal{G}$ under each choice of the subkeys of the last $r-7$ rounds. Therefore, the time complexity of our attack is $O({2^{(r - 7)n/4 + 2n/3}})$.

\noindent\textbf{Pre-computation phase.} As same as the computing process in Sect. \ref{pre-computation}, we construct the table $T_\delta$ classically without quantum computation. It will cost $O(2^{5n/6})$ time and $O(2^{5n/6})$ classical memory.

\noindent\textbf{Queried-data analysis phase.} We choose the same paired plaintexts in the Sect. \ref{query_phase}, and query for their ciphertexts, which are also stored in the table $T_{PC-all}$ indexed by $I$ ($0 \le I \le {2^{7n/6}} - 1$). After querying, we do following operations.
\begin{enumerate}
	
	\item Prepare the initial superposition state
	\begin{equation}\label{eq_r_initial}
		\left| {{\varphi _1}} \right\rangle  = \sum\limits_{K = 0}^{{2^{(r - 7)n/2}}} {{1 \over {\sqrt {{2^{(r - 7)n/2}}} }}{{\left| K \right\rangle }_1}} \sum\limits_{I = 0}^{{2^{7n/6}} - 1} {{1 \over {\sqrt {{2^{7n/6}}} }}} {\left| I \right\rangle _2},
	\end{equation}
	where $K = ({k_7},{k_8}, \ldots ,{k_{r - 1}})$.
	
	\item Query the paired plaintexts and ciphertexts from the table $T_{PC-all}$ by QRAM,
	\begin{equation}\label{eq_r_qram}
		\left| {{\varphi _2}} \right\rangle  = \sum\limits_{K = 0}^{{2^{(r - 7)n/2}}} {{1 \over {\sqrt {{2^{(r - 7)n/2}}} }}{{\left| K \right\rangle }_1}} \sum\limits_{I = 0}^{{2^{7n/6}} - 1} {{1 \over {\sqrt {{2^{7n/6}}} }}} {\left| I \right\rangle _2}{\left| {PC-all} \right\rangle _3}.
	\end{equation}
	
	\item Decrypt the ciphertexts partially to the values at ${v_7}||{v_6}$ for each $K$,
	\begin{equation}\label{eq_r_decrypt}
		\left| {{\varphi _3}} \right\rangle  = \sum\limits_{K = 0}^{{2^{(r - 7)n/2}}} {{1 \over {\sqrt {{2^{(r - 7)n/2}}} }}{{\left| K \right\rangle }_1}} \sum\limits_{I = 0}^{{2^{7n/6}} - 1} {{1 \over {\sqrt {{2^{7n/6}}} }}} {\left| I \right\rangle _2}{\left| {PC-all} \right\rangle _3}{\left| {{C_{{v_7}||{v_6}}}} \right\rangle _4},
	\end{equation}
	where ${C_{{v_7}||{v_6}}}$ represents the values at ${v_7}||{v_6}$.
	
	\item Do Grover search for that the difference $X'$ of the corresponding ${C_{{v_7}||{v_6}}}$ is all zero in the first $n/6$ bits, with $O(2^{n/12})$ time. This will increase the time complexity of step 2 in Algorithm \ref{algorithm_claw_finding_F_G} with $O({2^{n/3}}) = O({2^{n/4}} \times {2^{n/12}})$ time, but does not have an influence of the entire time complexity.
	\begin{equation}\label{eq_r_filter}
		\begin{split}
			\left| {{\varphi _4}} \right\rangle  =& \sum\limits_{K = 0}^{{2^{(r - 7)n/2}}} {{1 \over {\sqrt {{2^{(r - 7)n/2}}} }}{{\left| K \right\rangle }_1}} \otimes \\
			&\sum\limits_{0 \le I' \le {2^{7n/6}} - 1,\left| {I'} \right| = {2^n}} {{1 \over {\sqrt {{2^n}} }}{{\left| {I'} \right\rangle }_2}{{\left| {PC - al{l^{I'}}} \right\rangle }_3}{{\left| {C_{{v_7}||{v_6}}^{I'}} \right\rangle }_4}} 
		\end{split}
	\end{equation}

	\item Do outer Grover search for $K$ satisfying $\mathcal{H}(K) = 1$ and apply Algorithm \ref{algorithm_claw_finding_F_G} in the inner for each $K$. We define the function $\mathcal{H}:K \to \{ 0,1\} $. $\mathcal{H}(K) = 1$ if and only if there is claw after Algorithm \ref{algorithm_claw_finding_F_G} for the corresponding $K$, otherwise $\mathcal{H}(K) = 0$. 
\end{enumerate}

Note that $p_1=2^{-n/2}$, $p_2=2^{-2n/3}$, $N_c=2^{n/2}$, hence, $\delta$ = 4 is sufficient to filter out all the wrong candidates.

\section{Conclusion}\label{sec_conclusion}
To reduce the time complexity of classic and quantum attacks on $r$-round ($r \ge 7$) Feistel construction, we propose a new quantum meet-in-the-middle attack in Q1 model. We introduce a quantum claw finding algorithm and Grover's algorithm in Guo \textit{et al.}'s meet-in-the-middle attack to reduce complexity. Its time complexity only needs $O({2^{2n/3 + (r - 7)n/4}})$ and is less than classical and quantum attacks. Moreover, it belongs to the Q1 model and is more practical than other quantum attacks.

Furthermore, we hope to carry out quantum meet-in-the-middle attacks on more multi-rounds Feistel constructions. Because there are not only 5-round distinguisher, but also 7-round distinguisher, 8-round distinguisher, etc. Combining these distinguishers with quantum claw finding algorithms, or even other quantum algorithms may achieve good attack results.

\subsubsection{Acknowledgements}
This work was supported by the Open Fund of Advanced Cryptography and System Security Key Laboratory of Sichuan Province (Grant No. SKLACSS-202103) and the 13th Five-Year Plan National Cryptography Development Fund (MMJJ20180217).

%
%
%
\bibliographystyle{splncs04}
\bibliography{bibliography.bib}

\begin{thebibliography}{10}
\providecommand{\url}[1]{\texttt{#1}}
\providecommand{\urlprefix}{URL }
\providecommand{\doi}[1]{https://doi.org/#1}

\bibitem{Aoki2009}
Aoki, K., Guo, J., Matusiewicz, K., Sasaki, Y., Wang, L.: Preimages for
  step-reduced sha-2. In: Matsui, M. (ed.) Advances in Cryptology -- ASIACRYPT
  2009. pp. 578--597. Springer Berlin Heidelberg, Berlin, Heidelberg (2009)

\bibitem{Bernstein1997}
Bernstein, E., Vazirani, U.: Quantum complexity theory. SIAM Journal on
  Computing  \textbf{26}(5), ~1411 (1997)

\bibitem{Bonnetain2020}
Bonnetain, X., Naya-Plasencia, M., Schrottenloher, A.: On quantum slide
  attacks. In: Paterson, K.G., Stebila, D. (eds.) Selected Areas in
  Cryptography -- SAC 2019. pp. 492--519. Springer International Publishing,
  Cham (2020)

\bibitem{Buhrman2005}
Buhrman, H., Dürr, C., Heiligman, M., Høyer, P., Magniez, F., Santha, M.,
  de~Wolf, R.: Quantum algorithms for element distinctness. SIAM Journal on
  Computing  \textbf{34}(6),  1324--1330 (2005).
  \doi{10.1137/S0097539702402780}

\bibitem{Coppersmith1994}
Coppersmith, D.: The data encryption standard (des) and its strength against
  attacks. IBM Journal of Research and Development  \textbf{38}(3),  243--250
  (1994). \doi{10.1147/rd.383.0243}

\bibitem{Demirci2008}
Demirci, H., Sel{\c{c}}uk, A.A.: A meet-in-the-middle attack on 8-round aes.
  In: Nyberg, K. (ed.) Fast Software Encryption. pp. 116--126. Springer Berlin
  Heidelberg, Berlin, Heidelberg (2008)

\bibitem{Dong2020A}
Dong, X., Dong, B., Wang, X.: Quantum attacks on some feistel block ciphers.
  Designs, Codes and Cryptography  \textbf{88},  1--25 (2020)

\bibitem{Dong2019A}
Dong, X., Li, Z., Wang, X.: Quantum cryptanalysis on some generalized feistel
  schemes. Science China Information Sciences  \textbf{62}(2),  22501 (2019)

\bibitem{Dong2018A}
Dong, X., Wang, X.: Quantum key-recovery attack on feistel structures. Science
  China Information Sciences  \textbf{61}(10), ~1--7 (2018)

\bibitem{Feistel1975}
Feistel, H., Notz, W., Smith, J.: Some cryptographic techniques for
  machine-to-machine data communications. Proceedings of the IEEE
  \textbf{63}(11),  1545--1554 (1975). \doi{10.1109/PROC.1975.10005}

\bibitem{Giovannetti2008}
Giovannetti, V., Lloyd, S., Maccone, L.: Quantum random access memory. Phys.
  Rev. Lett.  \textbf{100},  160501 (Apr 2008).
  \doi{10.1103/PhysRevLett.100.160501},
  \url{https://link.aps.org/doi/10.1103/PhysRevLett.100.160501}

\bibitem{Grover1997}
Grover, L.K.: Quantum computers can search arbitrarily large databases by a
  single query. Physical review letters  \textbf{79}(23), ~4709 (1997)

\bibitem{Guo2016}
Guo, J., Jean, J., Nikoli{\'c}, I., Sasaki, Y.: Extended meet-in-the-middle
  attacks on some feistel constructions. Designs, Codes and Cryptography
  \textbf{80}(3),  587--618 (2016)

\bibitem{Hosoyamada2018A}
Hosoyamada, A., Sasaki, Y.: Quantum demiric-sel{\c{c}}uk meet-in-the-middle
  attacks: applications to 6-round generic feistel constructions. In:
  International Conference on Security and Cryptography for Networks. pp.
  386--403. Springer (2018)

\bibitem{Isobe2013A}
Isobe, T., Shibutani, K.: All subkeys recovery attack on block ciphers:
  Extending meet-in-the-middle approach. In: Knudsen, L.R., Wu, H. (eds.)
  Selected Areas in Cryptography. pp. 202--221. Springer Berlin Heidelberg,
  Berlin, Heidelberg (2013)

\bibitem{Isobe2013B}
Isobe, T., Shibutani, K.: Generic key recovery attack on feistel scheme. In:
  Sako, K., Sarkar, P. (eds.) Advances in Cryptology - ASIACRYPT 2013. pp.
  464--485. Springer Berlin Heidelberg, Berlin, Heidelberg (2013)

\bibitem{ISO2010}
ISO/IEC: Information technology -- security techniques -- encryption algorithms
  -- part 3: block ciphers  (2010)

\bibitem{Ito2019}
Ito, G., Hosoyamada, A., Matsumoto, R., Sasaki, Y., Iwata, T.: Quantum
  chosen-ciphertext attacks against feistel ciphers. In: Matsui, M. (ed.)
  Topics in Cryptology -- CT-RSA 2019. pp. 391--411. Springer International
  Publishing, Cham (2019)

\bibitem{Kaplan2016B}
Kaplan, M., Leurent, G., Leverrier, A., Naya-Plasencia, M.: Breaking symmetric
  cryptosystems using quantum period finding. In: Crypto 2016-36th Annual
  International Cryptology Conference. pp. 207--237. Springer (2016)

\bibitem{Kaplan2016A}
Kaplan, M., Leurent, G., Leverrier, A., Naya-Plasencia, M.: Quantum
  differential and linear cryptanalysis. IACR Transactions on Symmetric
  Cryptology  \textbf{2016}(1),  71--94 (2016).
  \doi{10.13154/tosc.v2016.i1.71-94}

\bibitem{Kuwakado2010}
Kuwakado, H., Morii, M.: Quantum distinguisher between the 3-round feistel
  cipher and the random permutation. In: 2010 IEEE International Symposium on
  Information Theory. pp. 2682--2685. IEEE (2010)

\bibitem{Kuwakado2012}
Kuwakado, H., Morii, M.: Security on the quantum-type even-mansour cipher. In:
  2012 International Symposium on Information Theory and its Applications. pp.
  312--316. IEEE (2012)

\bibitem{Leander2017}
Leander, G., May, A.: Grover meets simon--quantumly attacking the
  fx-construction. In: International Conference on the Theory and Application
  of Cryptology and Information Security. pp. 161--178. Springer (2017)

\bibitem{Knudsen2002}
R., K.L.: The security of feistel ciphers with six rounds or less. Journal of
  Cryptology  \textbf{15}(3),  207--222 (2002).
  \doi{10.1007/s00145-002-9839-y},
  \url{https://doi.org/10.1007/s00145-002-9839-y}

\bibitem{Santoli2016}
Santoli, T., Schaffner, C.: Using simon's algorithm to attack symmetric-key
  cryptographic primitives. arXiv preprint arXiv:1603.07856  (2016)

\bibitem{Shi2018}
Shi, T., Jin, C., Guan, J.: Collision attacks against aez-prf for authenticated
  encryption aez. China Communications  \textbf{15}(2),  46--53 (2018)

\bibitem{Simon1997}
Simon, D.R.: On the power of quantum computation. SIAM journal on computing
  \textbf{26}(5),  1474--1483 (1997)

\bibitem{Todo2013}
Todo, Y.: Upper bounds for the security of several feistel networks. In: Boyd,
  C., Simpson, L. (eds.) Information Security and Privacy. pp. 302--317.
  Springer Berlin Heidelberg, Berlin, Heidelberg (2013)

\bibitem{Xie2019}
Xie, H., Yang, L.: Using bernstein--vazirani algorithm to attack block ciphers.
  Designs, Codes and Cryptography  \textbf{87}(5),  1161--1182 (2019)

\bibitem{Xu2021}
Xu, Y., Liu, W., Yu, W.: Quantum forgery attacks on copa, aes-copa and marble
  authenticated encryption algorithms. Quantum Information Processing
  \textbf{20}(4),  1--21 (2021)

\bibitem{Zhao2019}
Zhao, S., Duan, X., Deng, Y., Peng, Z., Zhu, J.: Improved meet-in-the-middle
  attacks on generic feistel constructions. IEEE Access  \textbf{7},
  34416--34424 (2019). \doi{10.1109/ACCESS.2019.2900765}

\end{thebibliography}

\end{document}